\documentclass[final,authoryear,1p,times]{elsarticle}

\usepackage{graphicx}

\usepackage{amssymb}

\journal{Advances in Space Research}

\begin{document}

\begin{frontmatter}



\title{\textbf{Stratification-induced scale splitting in convection}}

                                                                   	
\author[shch]{O. V. Shcheritsa\corref{cor}}
\address[shch]{Keldysh Institute of Applied Mathematics, Moscow, 125047 Russia}
\ead{shchery@mail.ru}

\author[avg]{A. V. Getling}
\address[avg]{Skobeltsyn Institute of Nuclear Physics, Lomonosov Moscow State
University,\\Moscow, 119991 Russia} \cortext[cor]{Corresponding
author} \ead{A.Getling@mail.ru}

\author[shch]{O. S. Mazhorova}
\ead{olgamazhor@mail.ru}

\begin{abstract}
The coexistence of motions on various scales is a remarkable
feature of solar convection, which should be taken into account
in analyses of the dynamics of magnetic fields. Therefore, it
is important to investigate the factors responsible for the
observed multiscale structure of solar convection. In this
study, an attempt is made to understand how the scales of
convective motions are affected by the particularities of the
static temperature stratification of a fluid layer. To this
end, simple models are considered. The equations of
two-dimensional thermal convection are solved numerically for a
plane horizontal fluid layer heated from below, in an extended
Boussinesq approximation that admits thermal-diffusivity
variations. These variations specify the stratification of the
layer. The static temperature gradient in a thin sublayer near
the upper surface of the layer is assumed to be many times
larger than in the remainder of the layer. In some cases,
distributed heat sinks are assumed to produce a stably
stratified region overlying the convective layer.
Manifestations of the scale-splitting effect are noted, which
depend on the boundary conditions and stratification; {it
becomes more pronounced with the increase of the Rayleigh
number.} Small-scale convection cells are advected by
larger-scale flows. In particular, the phase trajectories of
fluid particles indicate the presence of complex attractors,
which reflect the multiscale structure of the flow. The effect
of the stably stratified upper sublayer on the flow scales is
also considered.
\end{abstract}


\end{frontmatter}


\section{Introduction}
\label{intr}

The dynamics of solar magnetic fields depends crucially on the
structure of the velocity field in the convection zone. In
particular, the coexistence of convective motions on various
scales is clearly reflected by the magnetic-field structure on
the photospheric levels.

At least four types of cellular structures, strongly differing
in their scale, can be identified with certainty on the solar
surface and attributed to the phenomenon of thermal convection,
viz., granules, mesogranules, supergranules and giant cells.
This multiscale structure is an important feature of solar
convection, which should be taken into account in studying the
dynamics of magnetic fields. It has not yet received a
convincing explanation, and an adequate hydrodynamic
description must be given to both the spatial structure of the
flows and the factors responsible for its development.

It can naturally be expected that, if convection cells are not
large in their plan size compared to the full thickness of the
convecting fluid layer, they should also be vertically
localised {(``suspended'')} in a relatively thin portion of
this layer. This is obviously possible if a certain sublayer
(height interval) with a convectively unstable thermal
stratification is contiguous with another sublayer where the
stratification is stable and exerts a braking action on the
convective motion. In this case, the flow nevertheless
penetrates into the stable region --- penetrative convection
occurs. If, however, the entire layer (from top to bottom) is
convectively unstable, the possibility for motion being
localised in a relatively thin sublayer is not so trivial.
{Such localisation can most easily be achieved by introducing,
e.g., sufficiently strong temperature variations of viscosity
\citep[see, e.g.,][]{stengetal}. At the same time, similar
effects of other factors remain little explored.} Even less
obvious is the coexistence of small cells with larger ones,
filling the entire layer thickness. {Such coexistence will here
be referred to as the scale-splitting effect.}

Quite likely, scale splitting could result from sharp changes
in the static vertical entropy gradient (or the static
temperature gradient in the case of an incompressible fluid) at
some heights. Under the conditions of the solar convection
zone, there are some prerequisites for this effect,
traditionally attributed to the enhanced instability of the
sublayers of partial ionisation of hydrogen and helium. {In
these sublayers, the specific heat of the plasma is increased
and the adiabatic temperature gradient accordingly decreased,
while the radiative gradient is increased. This idea was put
forward by \citet{simonleighton} in the context of the observed
coexistence of granulation and supergranulation and then
extended to mesogranulation by \citet{novetal}. It is typically
assumed that the zones of partial ionisation of neutral
hydrogen, neutral helium and singly ionised helium control the
formation of granules, mesogranules and supergranules,
respectively. The depths of location of these zones are
suggested to determine the characteristic sizes of these three
sorts of convection cells. The latter conjecture is, however,
fairly arbitrary from the standpoint of the theory of
convection, as we shall see below.}

{On the other hand, stratification effects are not the sole
candidate for the scale-splitting mechanism. Solar convection
is violent, turbulent fluid motion with a complex spectrum, and
hydrodynamic instabilities of some larger-scale motion can
produce secondary, smaller-scale flows. Conversely,
\citet{brcattoomre} note possible self-organisation processes
in turbulence (inverse energy cascades), which can give rise to
coherent structures. \citet{catlenzweiss} attribute the
formation of mesogranules to the collective interaction between
the granules, leaving the effects of stratifications beyond the
scope of their study.}

{Although the structure of solar convection has long been
explored using numerical simulations, the scale-splitting
effect has not yet received a convincing explanation. In
particular, \citet{derosaetal} simulate convection in a thin
shell and note the presence of various scales in the velocity
field; however, the large-scale flows computed in that study
are associated with global processes, while the size of the
smaller cells is controlled by the thickness of the shell.
These cells are not "suspended" near the outer boundary of the
shell. \citet{kitkosetal} investigate the multiscale flow
dynamics of vortical structures but do not reveal cellular
structures.}

{Previously, \citet{Get76,Get80} considered linear problems on
convection in layers with near-surface jumps in the static
temperature gradient, from low values in the bulk to high
values in a boundary sublayer. Such a jump was regarded as a
model representation of the sharp jump in the entropy gradient
at depths of order 1~Mm below the solar photosphere. In the
framework of the model, indirect indications for scale
splitting were found. These expectations were substantiated in
part by nonlinear numerical computations \citep{Get2007}. It
proved, however, that the tendency for convection cells to fill
the whole layer thickness is very strong, and small-scale,
near-surface convective motions develop only if the jump is
very sharp and the high-gradient sublayer is very thin.}
\begin{figure}
\centering
\includegraphics[width=0.44\linewidth, bb=40 0 418 360, clip]{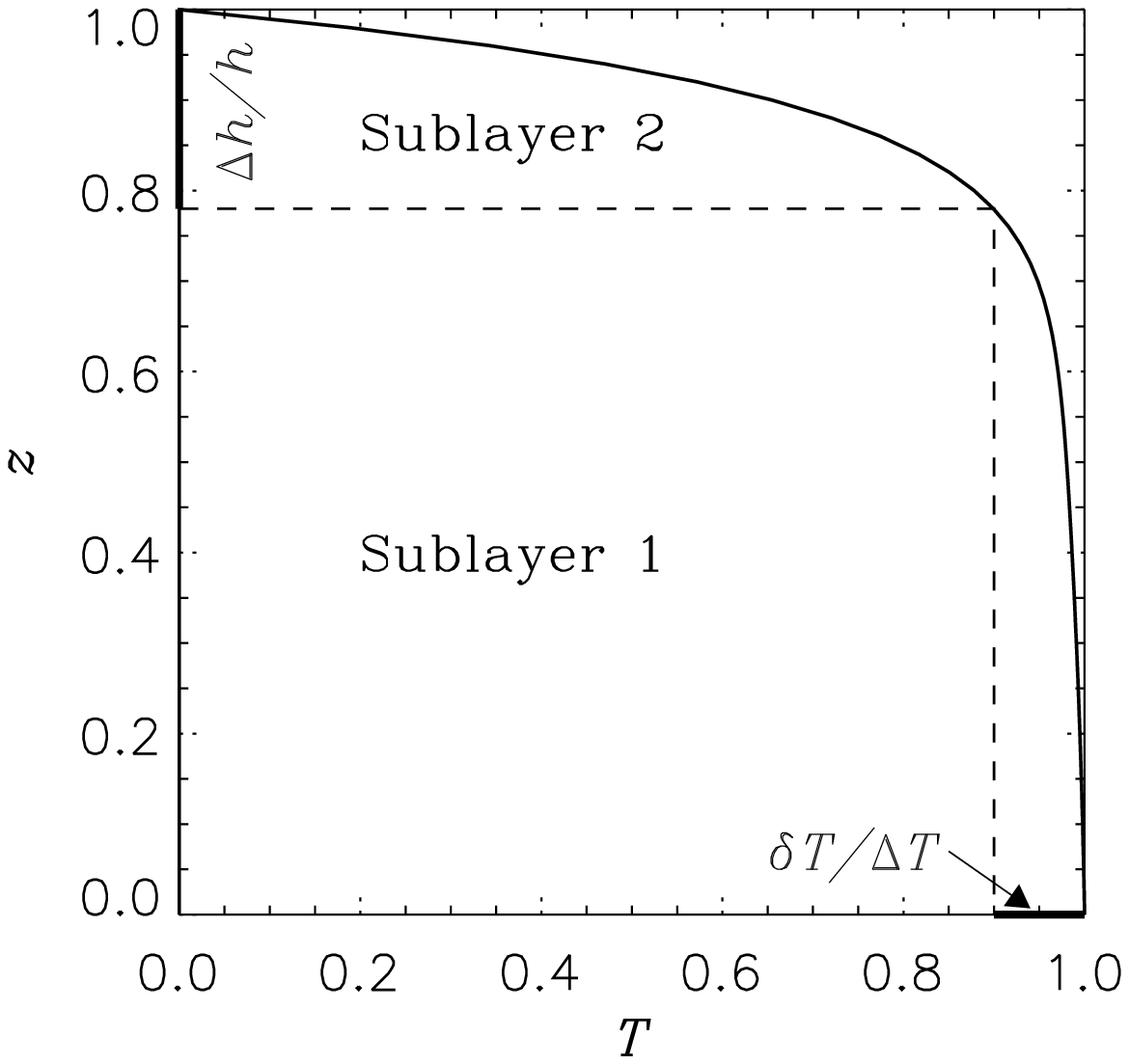}
\includegraphics[width=0.44\linewidth,bb=40 0 418 360, clip]{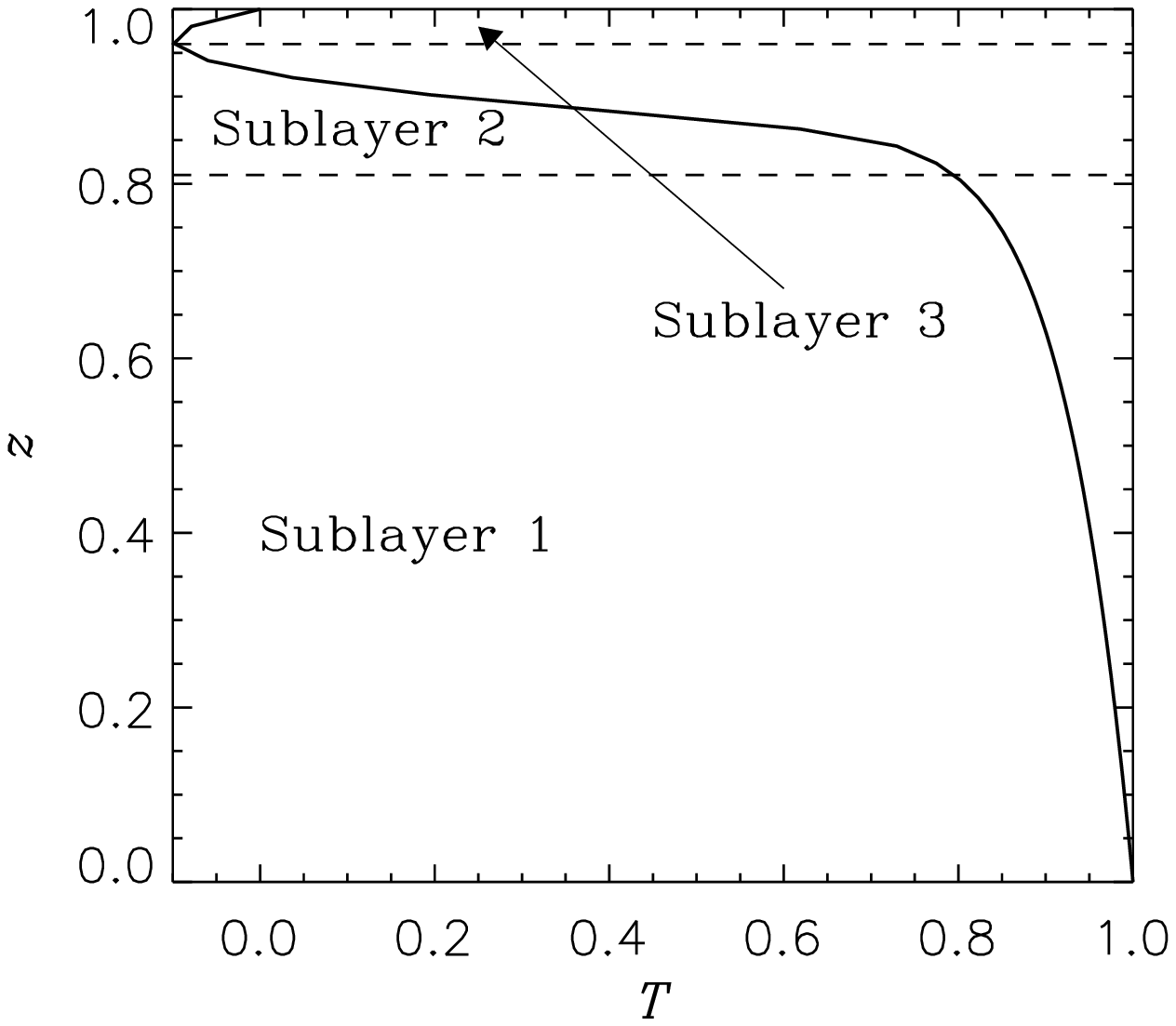}
{\phantom{.}\hspace{3.5cm} (a) \hspace{5.8cm} (b)}
  \caption{Static temperature profiles (in dimensionless variables):
(a)~monotonic profile for {$a = 5$, $b = 600$, $n = 10$};
(b)~nonmonotonic profile for $a=0.01$,  $b=600$, $n = 10$, $q_0=-2$, $z_0=0.8$
 }\label{stat_profil}
\end{figure}   

We investigate here possibilities of scale splitting by means
of numerically simulating two-dimensional convective flows in
the framework of similar simple models. {Although
two-dimensional simulations have only limited applicability to
natural fluid-dynamic systems, they are still of some interest,
being fairly simple and well tractable \citep{schmalzletal};
simulations of two-dimensional turbulent convection in a
density-stratified fluid layer by \citet{rogersetal} can be
mentioned as an example. Our simplified formulation of the
problem appears to be useful from the standpoint of evaluating
the role of some factors, taken alone, among those controlling
the static thermal stratification of the layer.}{\sloppy

}
\section{Formulation of the problem and numerical technique}

Assume that a plane horizontal layer $0<z<h$ of a viscous,
incompressible fluid is heated from below and consider its
finite segment $0<x<L$\ in which we shall simulate
two-dimensional ($\partial/\partial y=0$) convection flows. Let
the bottom and top boundaries be perfect thermal conductors and
let their temperatures be constant and equal to
$T_\mathrm{bot}=\Delta T>0$ and $T_\mathrm{top}=0$,
respectively. We also assume that the sidewalls of the region
are thermally insulated. The no-slip impermeability conditions
are specified at the bottom and side boundaries of the domain.
The top boundary may be either rigid (no-slip) or stress-free.

We are interested in situations where the static temperature
varies little (by $\delta T \ll \Delta T$) across the main
portion of the layer, of thickness $h - \Delta h$, $\Delta h
\ll h$  (Sublayer 1), while the most part of the temperature
difference, $\Delta T - \delta T$, corresponds to Sublayer~2
with a small thickness $\Delta h$, near the upper surface
(Fig.~\ref{stat_profil}).  {The kink near $z=h - \Delta h$ in
the temperature profile specified in this way qualitatively
resembles the transition (above depths of order 1~Mm) from the
bulk of the solar convection zone, where the stratification is
weekly superadiabatic, to the strongly unstable subphotospheric
layers.} To obtain such profiles, we assume the thermal
diffusivity to be temperature-dependent:
\begin{equation}
\label{chivsT}
\chi(T) = 1 + aT + bT^n.
\end{equation}
In some cases, we introduce heat sinks uniformly distributed
over the region above a certain level $z=z_0$ located in
Sublayer~2 to obtain temperature profiles with a minimum at a
certain height,  so that a stable Sublayer 3 be located above
Sublayer 2 (see below). This sublayer is considered to play the
role of a ``soft boundary,'' {or ``penetrable lid,''} as do the
stable layers located immediately above the temperature minimum
in the solar atmosphere. We shall give no attention to the flow
structure within this sublayer.

To describe the dynamics of convection, we use an extended
Boussinesq approximation, which admits thermal-diffusivity
variations \citep[for a discussion of different versions of the
Boussinesq  approximation, see, e.g.,][]{Getling_book}. If the
layer thickness $h$ is chosen as the unit length, $\Delta T$ as
the unit  temperature and the characteristic time of viscous
momentum transport, $\tau_\nu=h^2/\nu$, as the unit time ($\nu$
being the kinematic viscosity), the governing equations assume
the following dimensionless form:
\begin{eqnarray}
\label{NS}&&\frac{\partial {\bf v}}{\partial t}+({\bf v}%
\cdot\nabla){\bf v} =-\nabla\varpi+\mathbf{\hat z}\frac RP (T-T_\mathrm s)+\Delta {\bf v}, \\
\label{ht}&&\frac{\partial T}{\partial t} + {\mathbf v}\cdot\nabla T
= \frac 1P \nabla \cdot \frac{\chi(T)}{\chi(T_\mathrm{top})} \nabla T, \\[12pt]
\label{cnt}&&{}\nabla\cdot{\bf v}=0;
\end{eqnarray}
here, $\hat{\mathbf z}$ is the $z$--directed unit vector,
$T_\mathrm s(z)$ is the static temperature distribution,
$\chi(T)$ is the thermal diffusivity, $\varpi$ is the
nondimensionalised pressure and
$$
R=\frac{\alpha g\Delta T h^3}{\nu\chi(T_\mathrm{top})}\quad {\mathrm{and}} \quad
P=\frac{\nu}{\chi(T_\mathrm{top})}
$$
are the Rayleigh and Prandtl numbers, $\alpha$ being the volumetric coefficient
of thermal expansion of the fluid and $g$ the gravitational acceleration.

For two-dimensional incompressible flows, the stream function
$\psi$ and vorticity $\omega$ specified by the equations
$$
v_x=\frac{\partial\psi}{\partial z}, \quad
v_z=-\frac{\partial\psi}{\partial x}, \quad
\omega=-\Biggl(\frac {\partial^{2} \psi}{\partial x^{2}}+\frac
{\partial^{2}\psi}{\partial z^{2}} \Biggr)
$$
are variables convenient for constructing a computational
algorithm. We solve equations (\ref{NS})--(\ref{cnt}) with
properly chosen boundary conditions using the standard
procedure of splitting physical processes \citep{Yanenko81}.
Specifically, we first employ a matrix algorithm
\citep{Matrix_2,Matrix_REP} to determine the velocity field
from (\ref{NS}) written in terms of $\psi$ and $\omega$; next,
we find the temperature distribution in the layer from
(\ref{ht}). We use a conservative scheme of the second-order
accuracy in the spatial coordinates and of the first-order
accuracy in time \citep{Matrix_3}. Calculations are carried out
on a nonuniform grid, which is finer near the top and bottom
layer boundaries. The horizontal size of our computational
domain is $L=5\pi h = 15.7h$, and the total number of nodes is
$1024 \times 51$.

Most calculations were done for static temperature profiles
specified by dependences (\ref{chivsT}) with { $a = 5$ -- $20$,
$b = 600$, $n = 10$--$80$,} the temperature varying
monotonically in these cases (Fig.~\ref{stat_profil}a).
Alternatively, nonmonotonic profiles (Fig.~\ref{stat_profil}b)
were obtained by specifying heat sinks uniformly distributed
with a density $-q$ (i.e., {heat sources with a negative
density $q$}) above the height $z_0$, as the solution of the
problem
$$
\frac{\partial}{\partial
z}\left(\chi(T)\frac{\partial T}{\partial z}\right){+}q=0,\quad q=
\left\{
           \begin{array}{ll}
            q_0<0, &  z>z_0 \\
           0,   &  z<z_0 \\
           \end{array}
           \right. .
$$

{For the sake of comparisons, we also considered the case where
the static temperature profile of the form shown in
Fig.~\ref{stat_profil}a was produced by specially chosen
temperature-dependent heat sources, with a constant thermal
diffusivity (\ref{chivsT}). The results differ substantially
from those obtained for the case of the temperature-dependent
thermal diffusivity, which therefore appears to be physically
more appropriate in the context of solar convection.}

In our simulations, the flow is initiated by introducing random
thermal perturbations at a certain height within the upper
sublayer.
\begin{figure}[h+]
\includegraphics[width=0.5\textwidth,bb=5 90 190 220,clip]{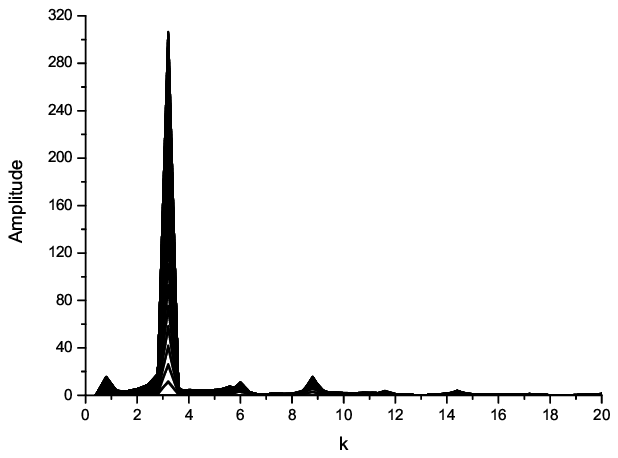}
\includegraphics[width=0.5\textwidth,bb=5 90 190 220,clip]{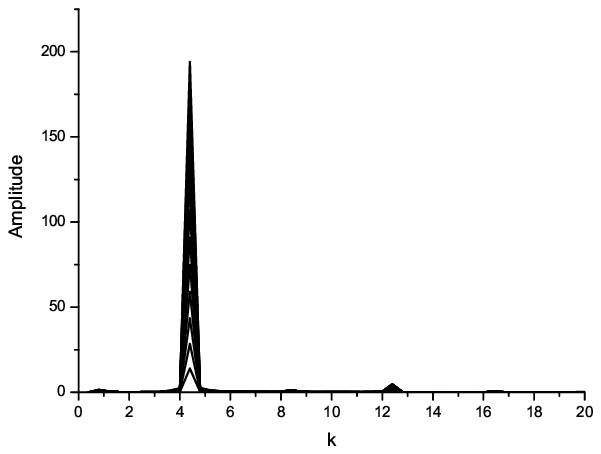}\\
\phantom{.}\hspace{3.45cm}(a)\hspace{6.35cm}(b)\\
\includegraphics[width=0.5\textwidth,bb=5 90 190 220,clip]{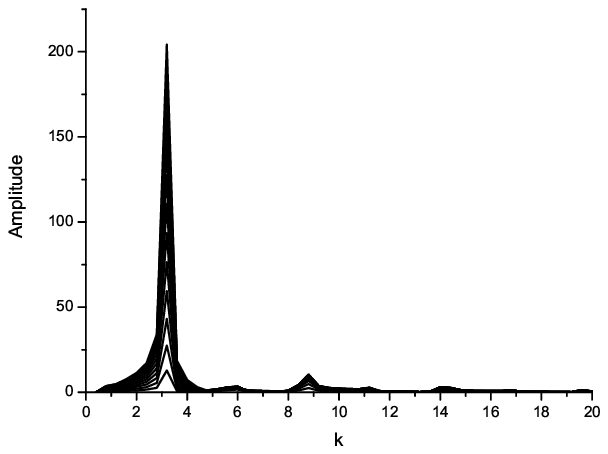}
\includegraphics[width=0.5\textwidth,bb=5 90 190 220,clip]{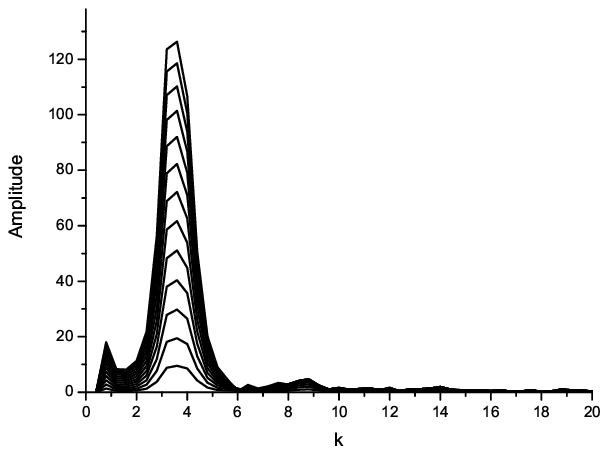}\\
\phantom{.}\hspace{3.45cm}(c)\hspace{6.35cm}(d)\\
\caption{Fourier spectra of simulated convective flows ($a=5$, $b=600$, $n=10$;
different curves in the same graph correspond to different heights $z$):
(a) rigid upper boundary, $R=10R_{\mathrm c}$ ($R_{\mathrm c}=1.95\times 10^6$);
(b) free upper boundary, $R=10R_{\mathrm c}$ ($R_{\mathrm c}=1.8\times 10^6$);
(c) rigid upper boundary, $R=55R_{\mathrm c}$ ($R_{\mathrm c}=1.95\times 10^6$);
(d) free upper boundary $R=55R_{\mathrm c}$ ($R_{\mathrm c}=1.8\times 10^6$).
}\label{Fourier_spectrum}
\end{figure}       

\section{Simulation results}
\subsection{Monotonic static temperature profiles}

{Some preliminary results of our simulations were reported by
\citet{gmshch}. Here, we substantially extend the explored
region of parameter space (in particular, to higher Rayleigh
numbers) and discuss the results more comprehensively. In
particular, along with the Fourier transform, we use the
technique of phase trajectories of fluid particles to analyse
the flow structure.}

For both the no-slip and stress-free boundary conditions at the
upper surface of the layer, the critical Rayleigh number
$R_\mathrm c$ was determined in the process of simulation. In
the regimes studied, { it lies in the range, roughly,
$R_\mathrm c = 1.95 \times 10^6$--$3.8 \times 10^6$ (depending
on $a$, $b$ and $n$)} for a rigid upper boundary and {
$R_\mathrm c = 4\times 10^5$--$1.8 \times 10^6$} for a free
upper boundary; the Prandtl number in our calculations is $P =
1$. {We present here our simulations for two degrees of
supercriticality, $R=10R_{\mathrm c}$ and $R=55.5R_{\mathrm
c}$.} The qualitative features of the results turn out to be
little sensitive to the choice of parameters $a$ and $n$ in the
above-mentioned ranges.

\begin{figure}
\centering
(a) \includegraphics [width=0.95\linewidth]{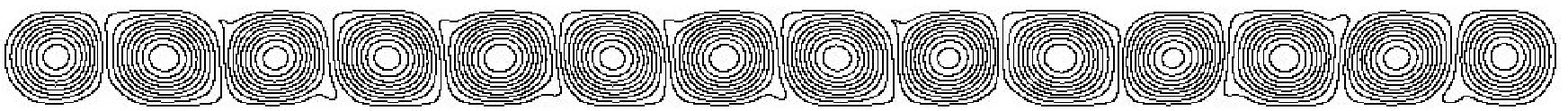}\\
(b) \includegraphics[width=0.95\linewidth]{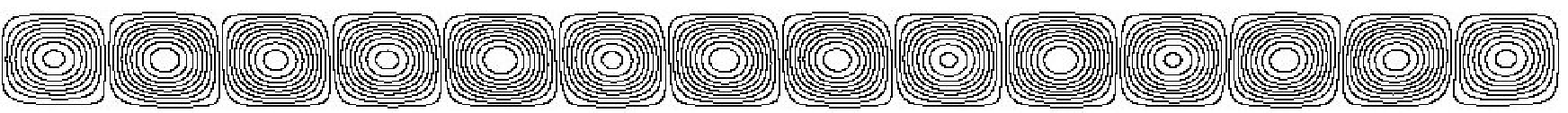}\\
(c) \includegraphics[width=0.95\linewidth]{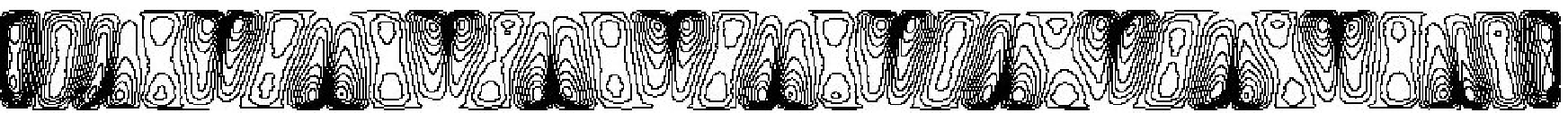}\\
\caption{Convective flow computed for a rigid upper boundary and
$R=10R_{\mathrm c}$ ($R_{\mathrm c}=1.95 \times 10^6$), $P = 1$, $a=5$, $b=600$, $n=10$,
with large cells and a smaller-scale flow coexisting: (a) streamlines (contours of the
stream function); (b) large-scale structures separated from the flow
using an ideal low-pass filter; (c) small-scale structures obtained by subtracting
the large-scale component from the full stream-function field. This flow pattern does not
undergo qualitative changes after $t\sim 0.1$, remaining weakly time-dependent
(we recall that the time is measured in the units of the time of viscous
momentum transport, $\tau_\nu$).
 }\label{techenie_rigid_new} 
\end{figure}

\begin{figure}
\centering
(a) \includegraphics [width=0.95\linewidth]{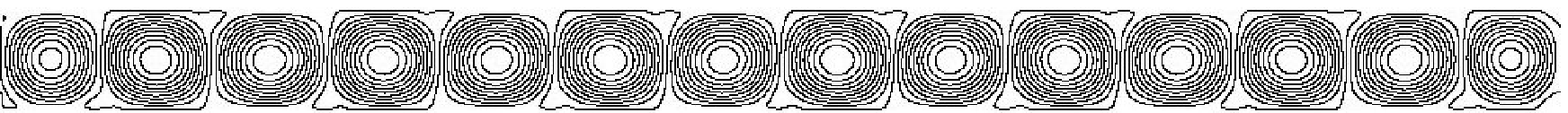}\\
(b) \includegraphics[width=0.95\linewidth]{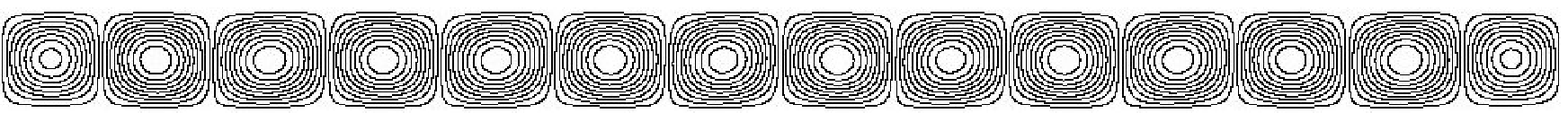}\\
(c) \includegraphics[width=0.95\linewidth]{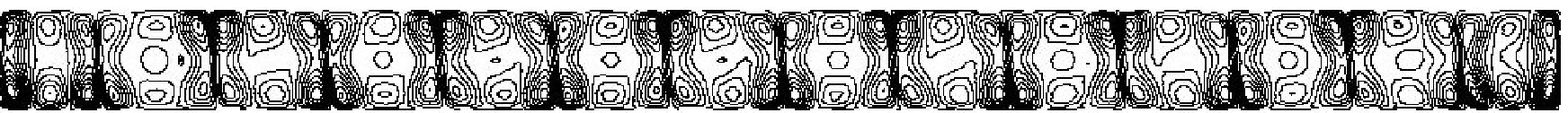}\\
 \caption{Same as in Fig.~\ref{techenie_rigid_new} but for $R=55.5R_{\mathrm c}$
 This pattern has been formed by $t\sim 0.05$ and does not undergo qualitative changes at
 later times, although is not quite stationary.}
 \label{techenie_rigid_R}
\end{figure}

We analyse the flow structure using the discrete Fourier
transform of the stream function with respect to the horizontal
coordinate $x$ at given heights $z$. High-frequency harmonics
are present in the spectrum, whose amplitudes amount to
{$5$--$16\%$ of the amplitude of the fundamental mode,
depending on the parameters of the problem
(Fig.~\ref{Fourier_spectrum})}. To separate the small-scale
component of the velocity field and visualise the fine cellular
structures present in the flow, we use an ideal low-pass filter
\citep{filters} and subtract the obtained large-scale component
from the original field. {This procedure proved to be efficient
in the case of the rigid upper boundary
(Figs~\ref{techenie_rigid_new}, \ref{techenie_rigid_R}). In the
case of the free upper boundary, however, the more complex
appearance of the spectra made our attempts of filtering
unsuccessful.}

If the upper horizontal boundary is rigid, motion starts
developing as small-scale convection in Sublayer~2. Later, the
disturbances penetrate deeper and gradually involve the entire
layer depth. As a result, large-scale convection rolls emerge,
with a width of about the layer thickness. As this takes place,
the small-scale flow in the upper sublayer does not disappear
and assumes the form of smaller  rolls with a size {typically
exceeding the sublayer thickness.} A flow of a similar small
scale also develops near the bottom layer boundary. {It should
be noted that, if the temperature profile is linear (the
classical Rayleigh--B\'enard problem), simulations at the same
$R$ and $P$ do not reveal small-scale structures.} {In
moderately supercritical regimes, the small rolls are
especially pronounced above and below the contact sections of
large rolls. As the Rayleigh number is increased, the small
rolls occupy a progressively wider region, and a tendency to
the formation of two layers of small-scale rolls is observed.}
Thus, in the fluid layer stratified on the whole unstably,
large convection cells\footnote{ By a convection cell in a
two-dimensional flow, a pair of neighbouring rolls is meant.}
filling the entire layer depth coexist with smaller ones
localised in relatively thin sublayers
(Figs~\ref{techenie_rigid_new}, \ref{techenie_rigid_R}). {The
scale-splitting effect is more pronounced at higher Rayleigh
numbers: the near-surface cells singled out by subtracting the
large-cell flow are smaller and more clear-cut at
$R=55.5R_{\mathrm c}$ (Fig.~\ref{techenie_rigid_R}) than at
$R=10R_{\mathrm c}$ (Fig.~\ref{techenie_rigid_new}). It} {is
worth noting that their location is not directly controlled by
the thickness sublayer of the large static temperature gradient
(Sublayer 2 in Fig.~\ref{stat_profil}a).}

\begin{figure}[h+]
\centering
(a) \includegraphics[width=0.95\linewidth]{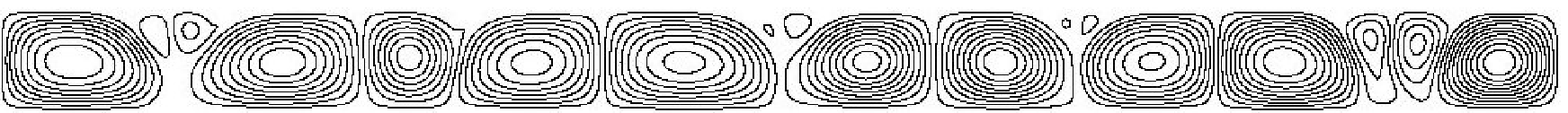}\\
(b) \includegraphics[width=0.95\linewidth]{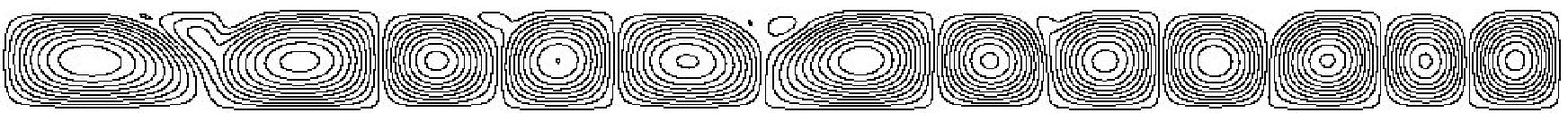}\\
(c) \includegraphics[width=0.95\linewidth]{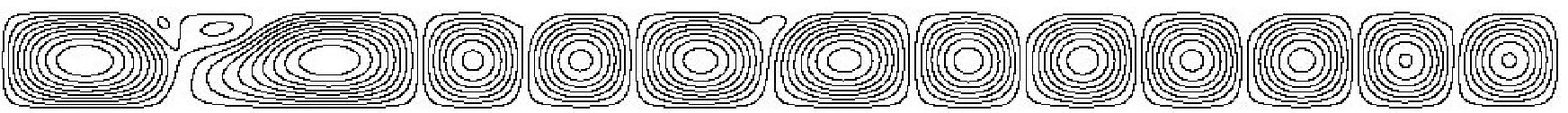}\\
(d) \includegraphics[width=0.95\linewidth]{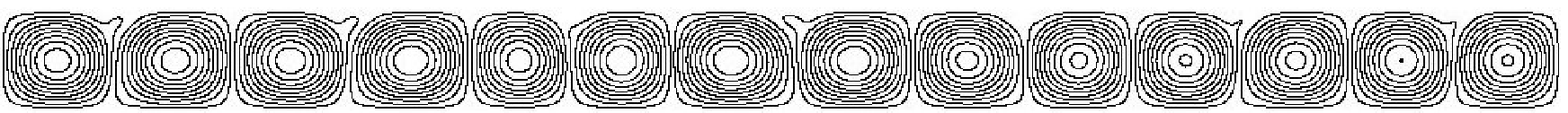}\\
  \caption{ Flow evolution in the case of a free upper boundary, at
  $R=10R_{\mathrm c}$ ($R_{\mathrm c}=1.8\times 10^6$), $P = 1$,  $a=5$, $b=600$, $n=10$.
  Streamlines (contours of the stream function) are shown for different times:
  (a) $t = 0.04$, small-scale cells originating near the upper surface of the layer;
  (b) $t = 0.045$, the ``breaking'' effect of the small cells, penetrating into the layer depth,
  on those large rolls whose horizontal size considerably exceeds the scale optimum;
  (c) $t = 0.14$, the flow structure formed in the process of penetration of small
  cells deeper into the layer;
  (d) $t = 0.22$, a tendency toward the expansion of large-scale cells.
  The emergence and development of small-scale structures in the upper sublayer.
  The flow pattern does not undergo qualitative changes at later times.
  }\label{techenie_free}
\end{figure}

\begin{figure}[h+]
\centering
(a) \includegraphics[width=0.95\linewidth]{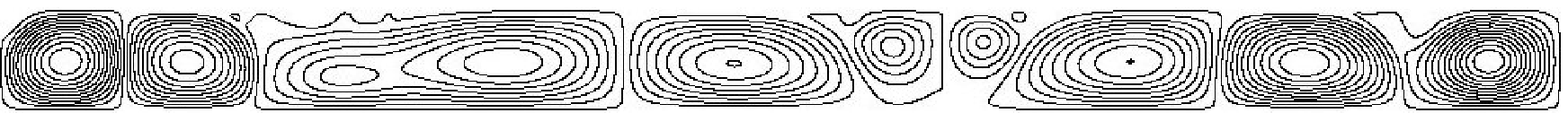}\\
(b) \includegraphics[width=0.95\linewidth]{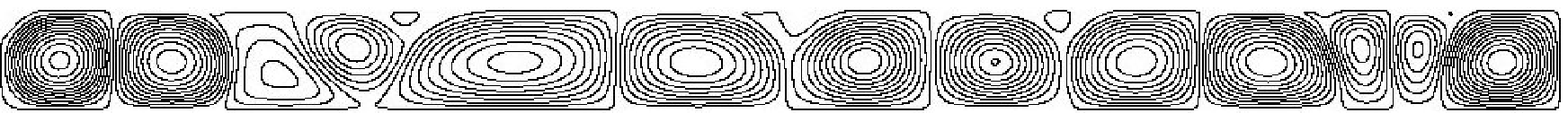}\\
(c) \includegraphics[width=0.95\linewidth]{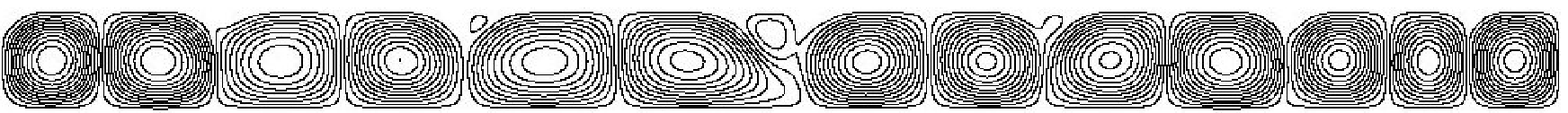}\\
(d) \includegraphics[width=0.95\linewidth]{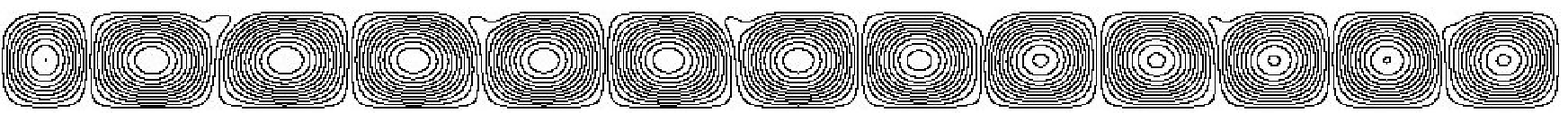}\\
  \caption{Same as in Fig.~\ref{techenie_free} but for
  $R=55R_{\mathrm c}$ and (a) $t = 0.006$, (b) $t = 0.007$,
  (c) $t = 0.008$ and (d) $t = 0.076$. The flow pattern does not undergo
  qualitative changes at later times.
  }\label{techenie_free_r}
\end{figure}

In the case of a free upper boundary, the flow also originates
in the upper sublayer, after which the emerged small-scale
structures penetrate deeper into the layer, stimulating the
formation of a flow throughout the layer. The cells in
Sublayer~1 tend to grow in size; however, the small-scale
structures present in Sublayer~2 control this process: as the
horizontal size of a large structure becomes considerably
larger than the layer thickness, the cells moving down from the
upper sublayer break this structure into two portions
(Figs~\ref{techenie_free}, \ref{techenie_free_r}). In contrast
to the case of a rigid upper boundary, where the number of
large rolls is constant, the number of large-scale structures
now varies between 10 and 15.

\begin{figure}[h!]
\centering
(a) \includegraphics[width=0.95\linewidth]{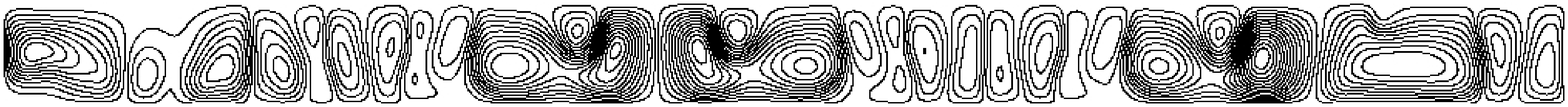}\\
(b) \includegraphics[width=0.95\linewidth]{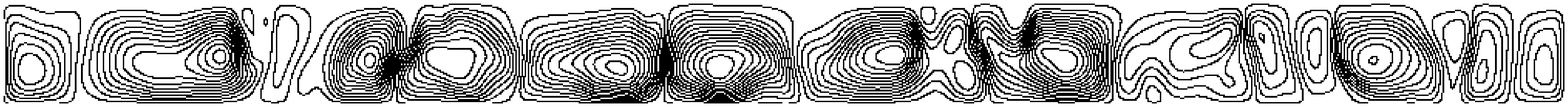}\\
(c) \includegraphics[width=0.95\linewidth]{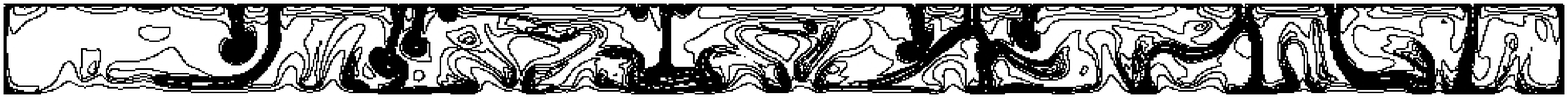}\\
(d) \includegraphics[width=0.95\linewidth]{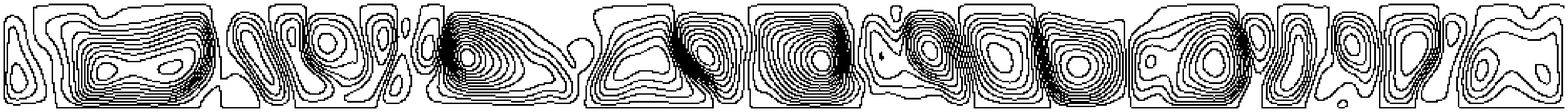}\\
(e) \includegraphics[width=0.95\linewidth]{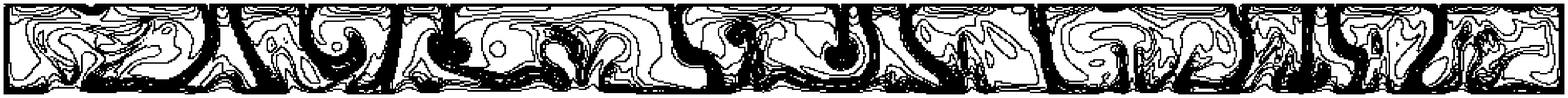}
  \caption{Flow structure in the case of the nonmonotonic static
  temperature profile, for $R=20\,000\approx 1.5R_{\mathrm c}$, $P = 1$
  and different times:
  (a) stream function at $t=0.21$,
  small-scale structures penetrate deep into the layer;
  (b) stream function at $t=0.42$,
  developing small-scale structures either glide deep into
  the layer or break a large structure into two parts;
  (c) temperature distribution at $t=0.42$
  with local overheat and underheat zones present in the bulk
  of the layer; (d) stream function at $t=0.7$,
  small-scale structures are transferred by the large ones;
  (e) temperature distribution at $t=0.7$
  with local overheat and underheat zones present in the bulk
  of the layer. {The flow pattern does not undergo
  qualitative changes at later times.}
 }\label{heat}
\end{figure}    
\begin{figure}[h!]
\centering
(a) \includegraphics[width=0.95\linewidth]{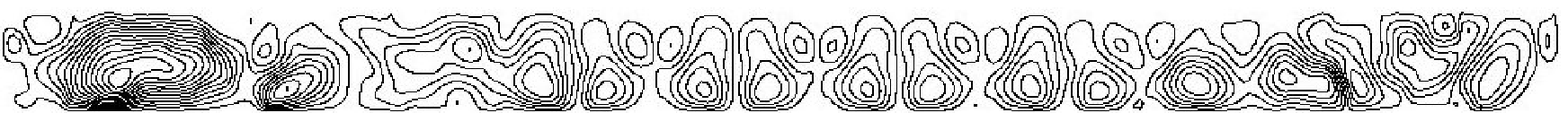}\\
(b) \includegraphics[width=0.95\linewidth]{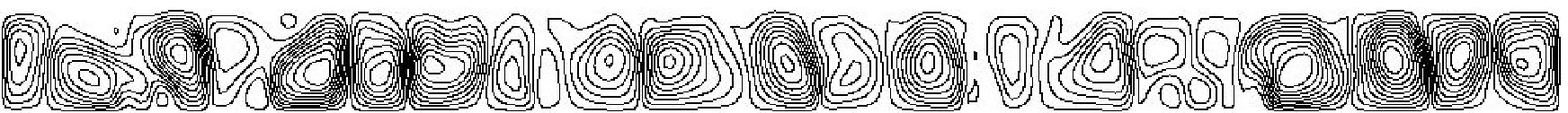}\\
(c) \includegraphics[width=0.95\linewidth]{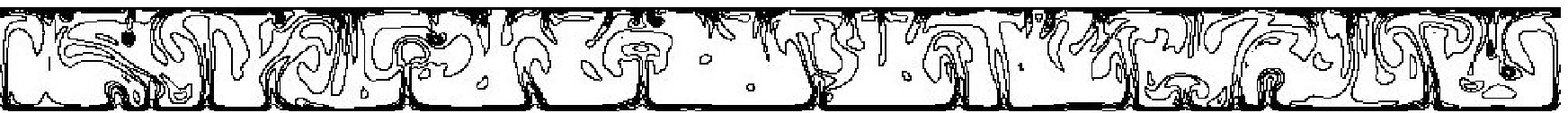}\\
(d) \includegraphics[width=0.95\linewidth]{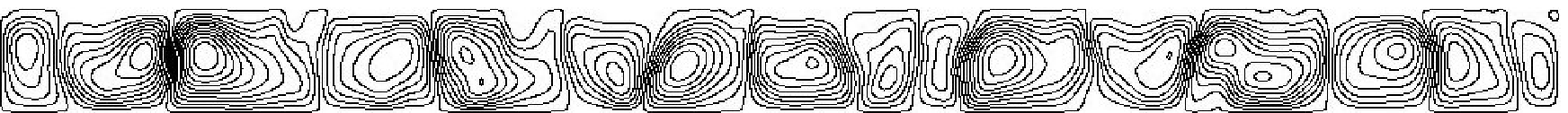}\\
(e) \includegraphics[width=0.95\linewidth]{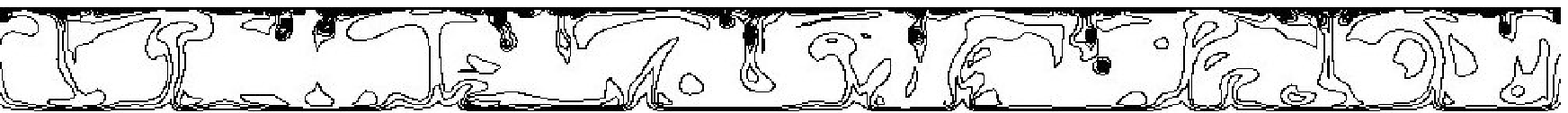}
  \caption{{Same as in Fig.~{heat} but for
  $R=200\,000\approx 15R_{\mathrm c}$, $P = 1$ and (a) $t=0.12$,
  (b, c) $t=0.32$, (d, e) $t=0.8$. The flow pattern does not
  undergo qualitative changes at later times.}
 }\label{heat_2}
\end{figure}    

The spectrum of the flow {(Figs~\ref{Fourier_spectrum}b, d)} is
now more complex than in the case of the rigid upper boundary
{(Figs~\ref{Fourier_spectrum}a, c)}. Since the number of large
structures is variable, the fundamental mode is also
time-dependent. Harmonics with different wavenumbers dominate
in the bulk of the layer and in the upper sublayer. The
enhancement of the small-scale component frequently parallels
with the weakening of the large-scale component, and vice versa
--- a sort of intermittency is observed. As this takes place,
the sublayers in which the cells are localised vary in their
thickness and do not coincide with the sublayers specified by
the static vertical temperature profile. In contrast to the
case of a rigid upper boundary, the small-scale structures are
localised only in the upper part of the layer.

{If the profile shown in Fig.~\ref{stat_profil}a is produced by
a heat-source distribution, the flow behaves in a very complex
manner and has a wide-band spectrum. Although the flow dynamics
deserves a careful analysis in this case, it does not appear to
be quite relevant to our subject. We plan to present such an
analysis elsewhere.}

\subsection{Nonmonotonic static temperature profile}

The nonmonotonic static temperature profile shown in
Fig.~\ref{stat_profil}b was created by combining a
temperature-dependent thermal diffusivity with distributed heat
sinks. The needed profiles were obtained at the following
parameters: $a =0.01$--$0.1$,  $b =600$, $n = 10$--$20$,
$q_0=-2$, $z_0=0.8$. In the regimes studied, the Rayleigh
number varied in the range {$R= 20\,000$--200\,000} and the
Prandtl number was $P =1$.

{The flows computed for the nonmonotonic profile are much less
ordered than in the case of the rigid or free upper boundary.
They are highly changeable, characterised by permanently}
{arising descending and ascending plumes, and can on the whole
be assigned to fairly developed turbulence.}

Convection was initiated by introducing a random perturbation
of the static temperature profile at a certain height. {The
dynamics of the flow is highly sensitive to where the initial
perturbations are introduced. For this reason, special efforts
are required to determine the critical Rayleigh number. At this
stage, our very crude estimates based on the simulated regimes
near the threshold of convective instability yielded $R_\mathrm
c \approx 13\,000$.}

Motion starts developing at the interface between Sublayers~2
and 3, after which small-scale structures penetrate to deeper
levels and stimulate the development of the flow in the bulk of
the layer (see Fig.~\ref{heat} {and especially
Fig.~\ref{heat_2})}. Large-scale structures form, with a
vertical size comparable with the layer thickness and
horizontal sizes exceeding it. The number of large structures
varies in the evolving flow pattern. Fresh small-scale
structures permanently originate at the interface between the
sublayers and either glide deeper between two large structures
or break a large structure into two parts, preventing it from
increasing its horizontal size (Figs \ref{heat}b,
{\ref{heat_2}b}). In the bulk of the layer, local overheat or
underheat zones emerge (see Figs \ref{heat}c, \ref{heat}e {and
also Figs \ref{heat_2}c, \ref{heat_2}e, where they are more
pronounced)}.

It can be seen from the Fourier transform of the stream
function {(not presented here)} that three or four modes
(harmonics) with incommensurate wavenumbers dominate in the
spectrum of a well-developed flow. Different modes dominate at
different times, so that determining the number of spatial
scales present is not a simple task. Applying
numerical-homology techniques to the velocity field
\citep{Homology_2} confirms the coexistence of several spatial
scales although does not enable us to determine the size and
localisation of the structures.

\subsection{Phase space and attractors}

The complex structures developing in various flows cannot
always be classified and analysed using Fourier-transform-based
methods. To make the presence of various flow scales more
apparent, we constructed the trajectories of fluid particles in
an appropriately defined phase space, $(z,v_z)$. Several
hundred particles differing in their initial position were
used. In their motion, all these particles repeatedly passed
from one structure to another, and transitions between large
structures and between a large and a small structure ---
possibly with a subsequent return to the large structure ---
could take place.

\begin{figure}[h+]
\includegraphics[width=0.5\textwidth,bb=42 15 504 330,clip]{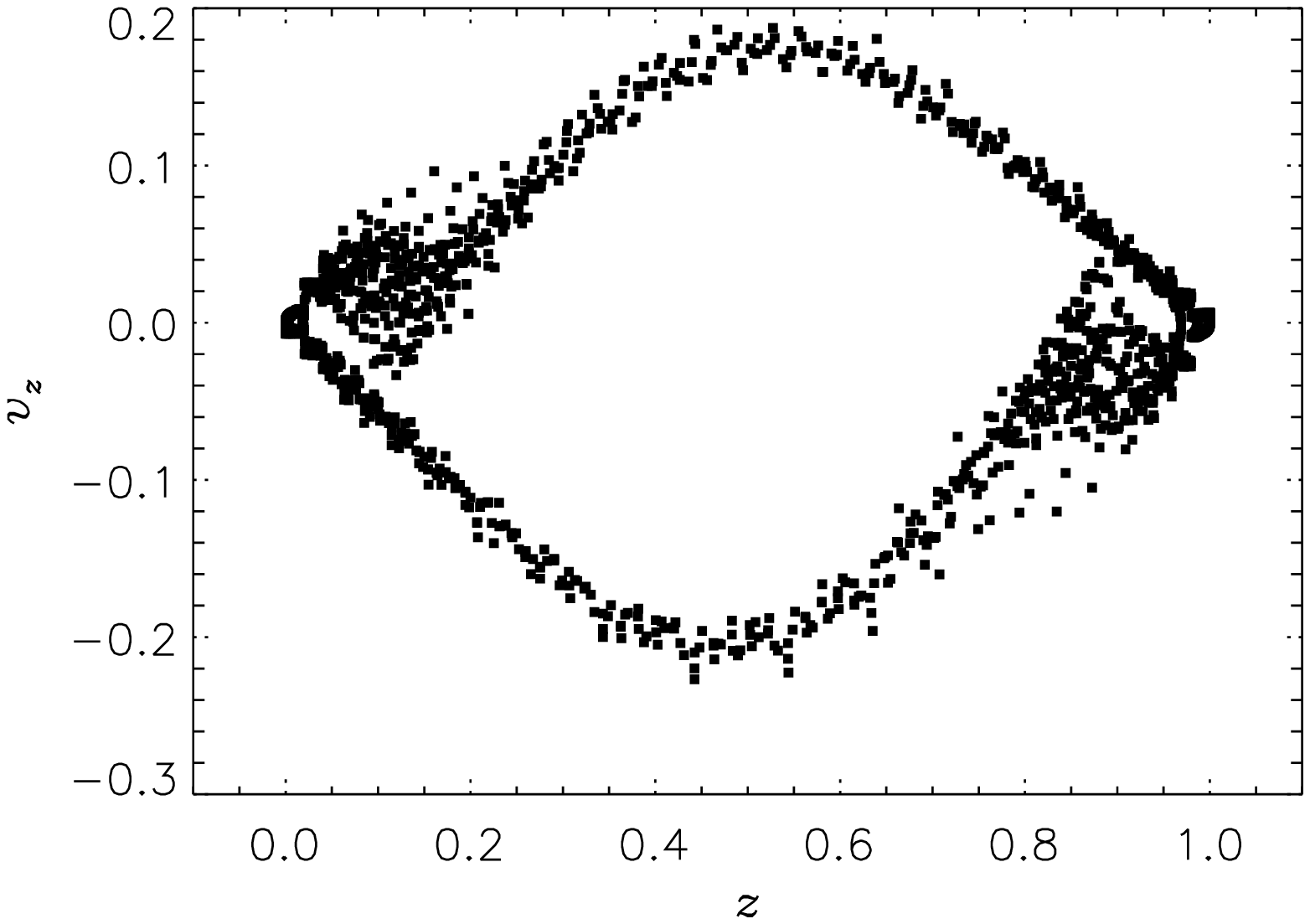}
\includegraphics[width=0.5\textwidth,bb=42 15 504 330,clip]{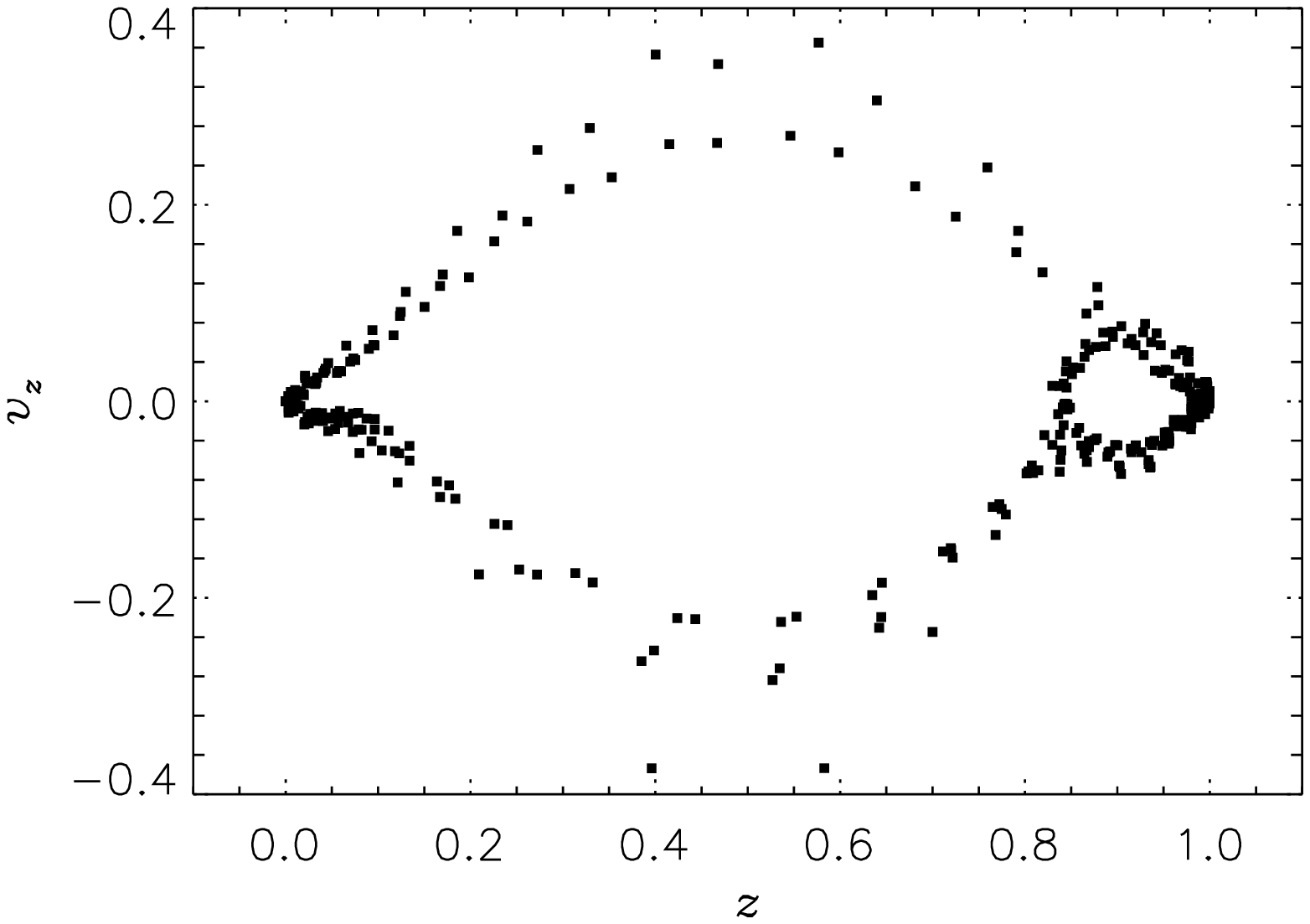}\\
\phantom{.}\hspace{3.3cm}(a)\hspace{6.4cm}(b)\\
\includegraphics[width=0.5\textwidth,bb=42 15 504 360,clip]{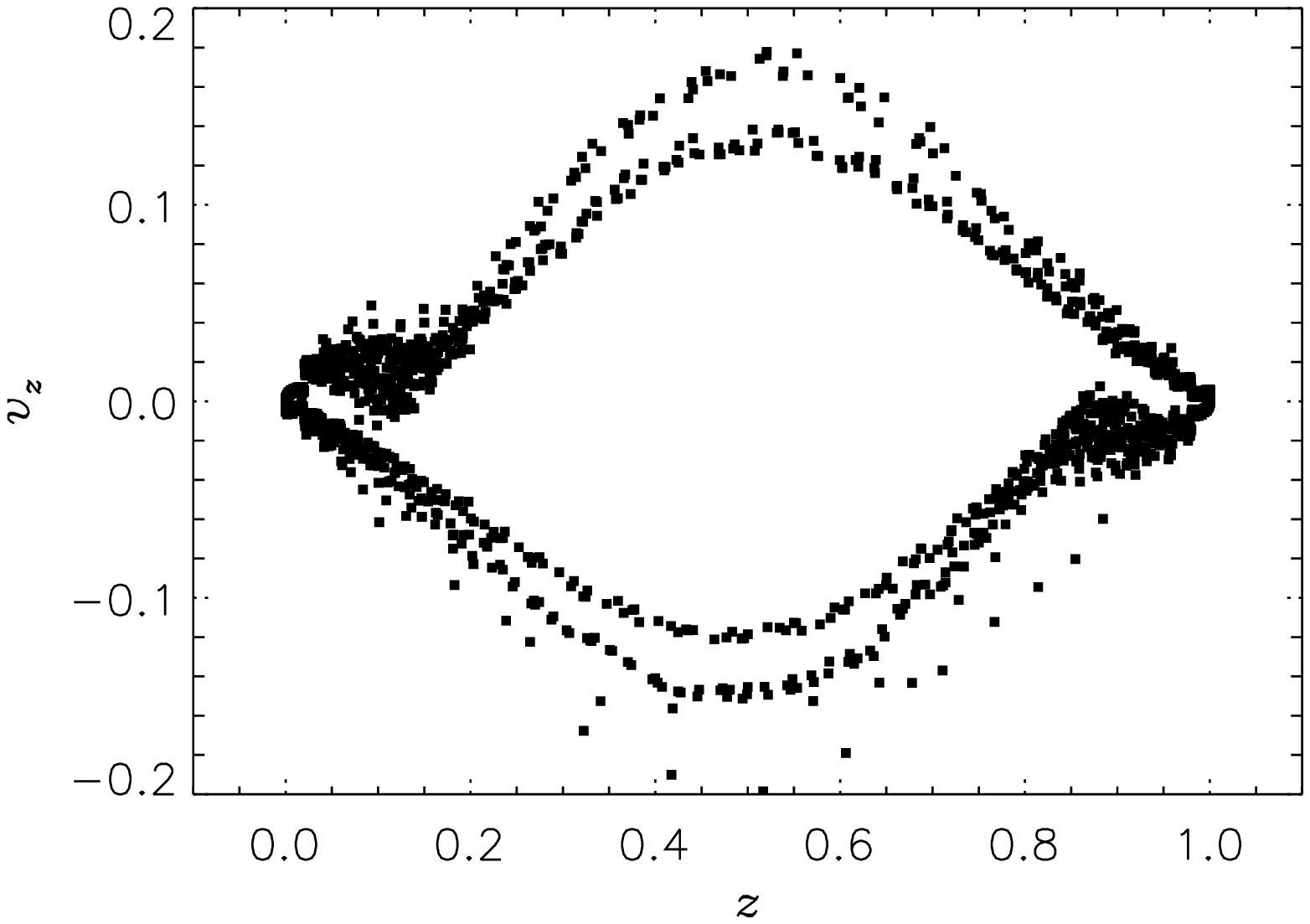}
\includegraphics[width=0.5\textwidth,bb=42 15 504 360,clip]{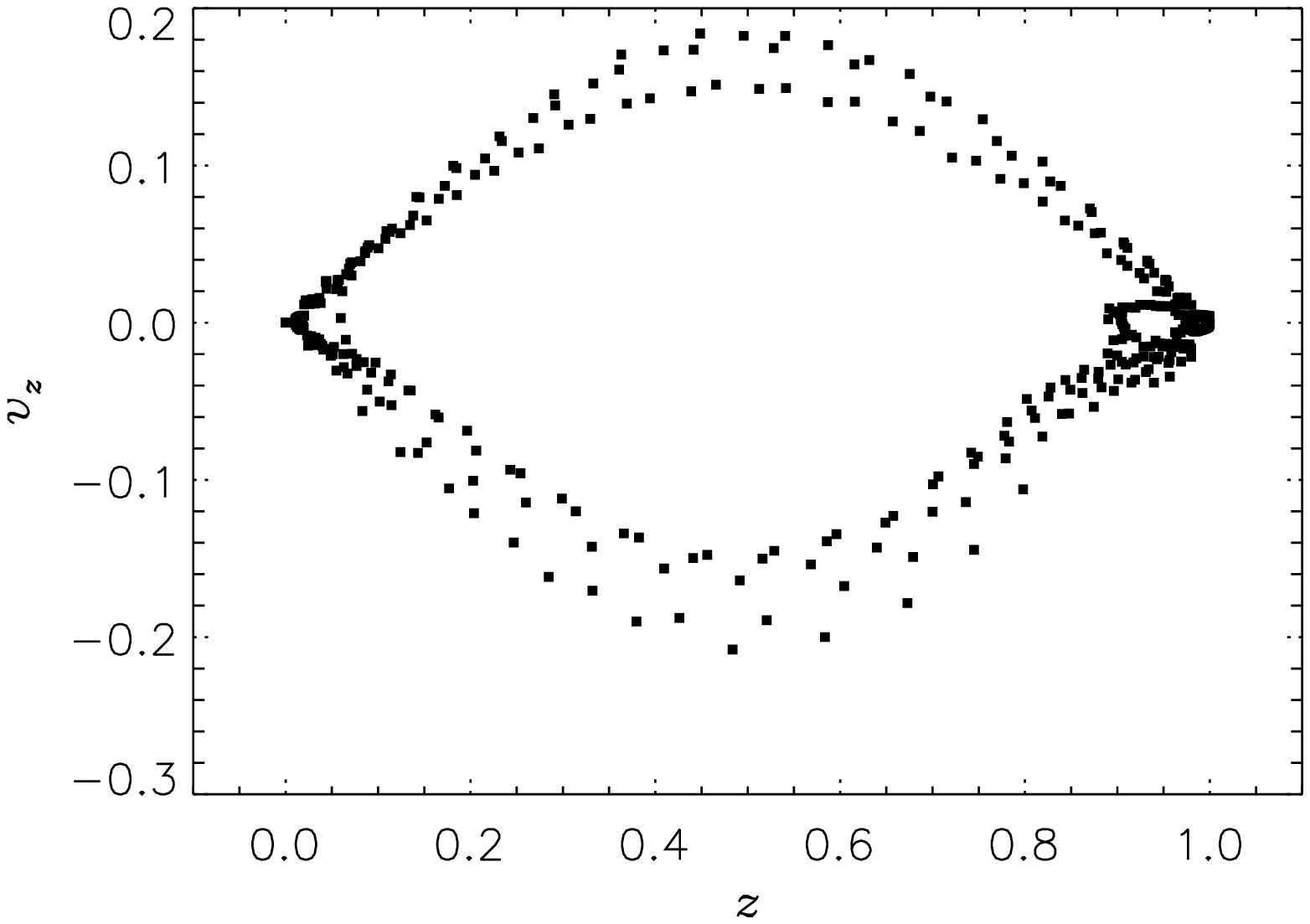}\\
\phantom{.}\hspace{3.3cm}(c)\hspace{6.4cm}(d)\\
\caption{ Phase trajectories of a particle in space $(z,v_z)$;
$a=5$, $b=600$, $n=1$:
(a) rigid upper boundary, $R=10R_{\mathrm c}$ ($R_{\mathrm c}=1.95\times10^6$);
(b) free upper boundary, $R=10R_{\mathrm c}$ ($R_{\mathrm c}=1.8\times10^6$),
(c) rigid upper boundary, $R=55R_{\mathrm c}$ ($R_{\mathrm c}=1.95\times10^6$);
(d) free upper boundary, $R=55R_{\mathrm c}$ ($R_{\mathrm c}=1.95\times10^6$).
}\label{attr_rigid}
\end{figure}

If the static temperature profile is monotonic, an attractor
can be detected whose structure clearly demonstrates the
presence of different flow scales. In Fig.~\ref{attr_rigid}a,
which refers to the case of a rigid upper boundary, the
trajectory of only one particle is shown, others being quite
similar. The velocity vanishes near the upper and lower
boundaries ($z=0$ and $z=1$) according to the no-slip
condition, and a passage of the trajectory from the positive to
negative half-plane reflects a transition from ascending to
descending motion. Therefore, the large ellipses correspond to
the large structure and the small ellipses near $z=0$ and $z=1$
represent the small-scale flows near the upper and lower
boundaries. If the upper boundary is free, small-scale motions
are localised only near this boundary (Fig.~\ref{attr_rigid}b).

{It can be found from Fig.~\ref{attr_rigid}a that, for
$R=10R_{\mathrm c}$ and the rigid upper boundary, the vertical
size of the large structures is equal to the layer thickness
$h$, the size of the small structure being} {$0.18h$ near the
lower boundary and $0.14h$ near the upper boundary. Different
scales of the small structures can be due to different thermal
diffusivities near $z=0$ and $z=1$:
$\chi(T_{\mathrm{bot}})=605$, while $\chi(T_{\mathrm{top}})=1$.
In the case of $R=55R_{\mathrm c}$, the size of the small
structures is $0.17h$ near the lower boundary and $0.11h$ near
the upper boundary (Fig. \ref{attr_rigid}c). If the upper
boundary is free, the vertical size of the small structures is
$0.18h$ at $R=10R_{\mathrm c}$ (Fig. \ref{attr_rigid}b) and
$0.11h$ at $R=55R_{\mathrm c}$ (Fig. \ref{attr_rigid}d).}

\begin{figure}[h+]
\includegraphics[width=0.5\textwidth,bb=20 15 504 340,clip]{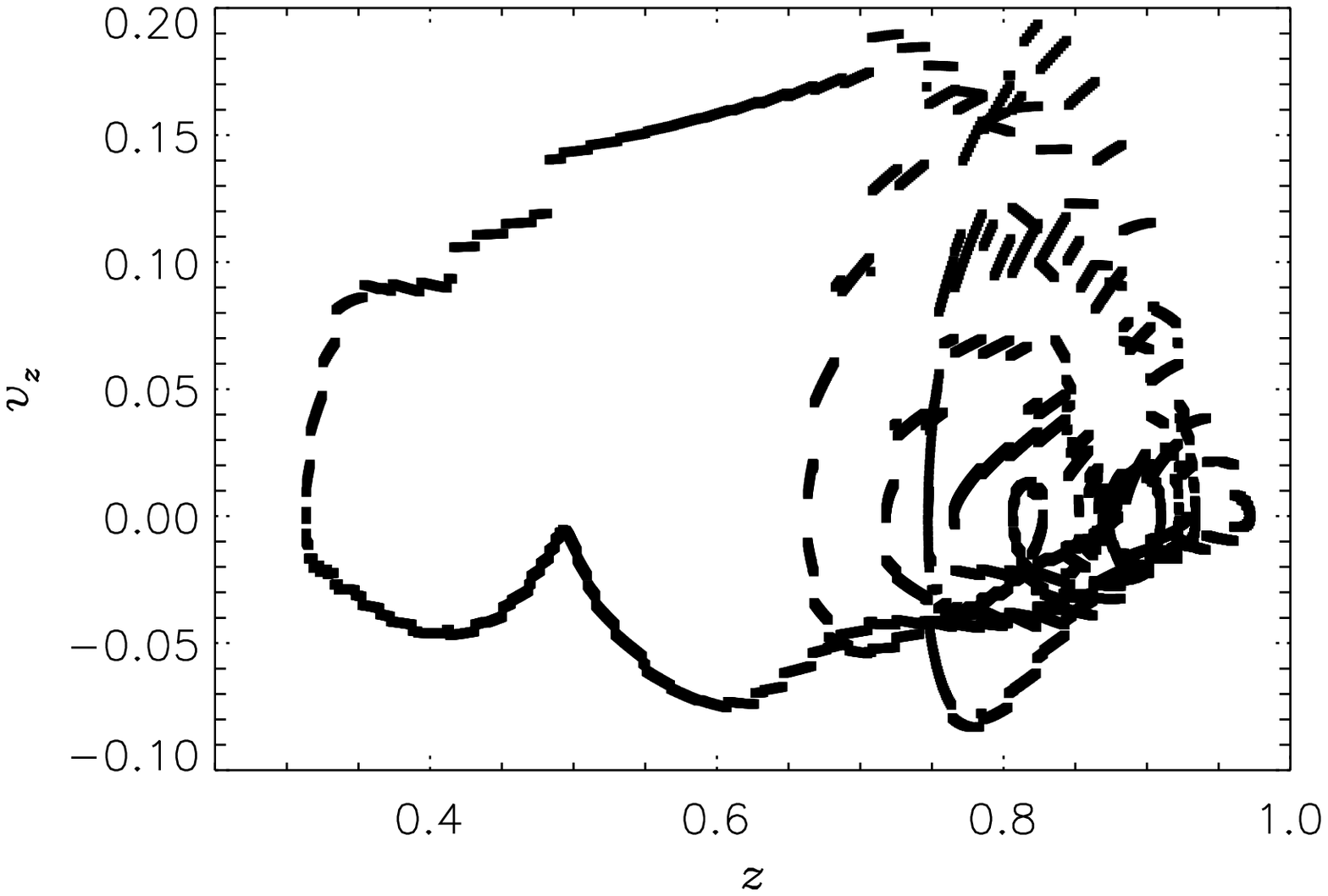}
\includegraphics[width=0.5\textwidth,bb=30 15 504 360,clip]{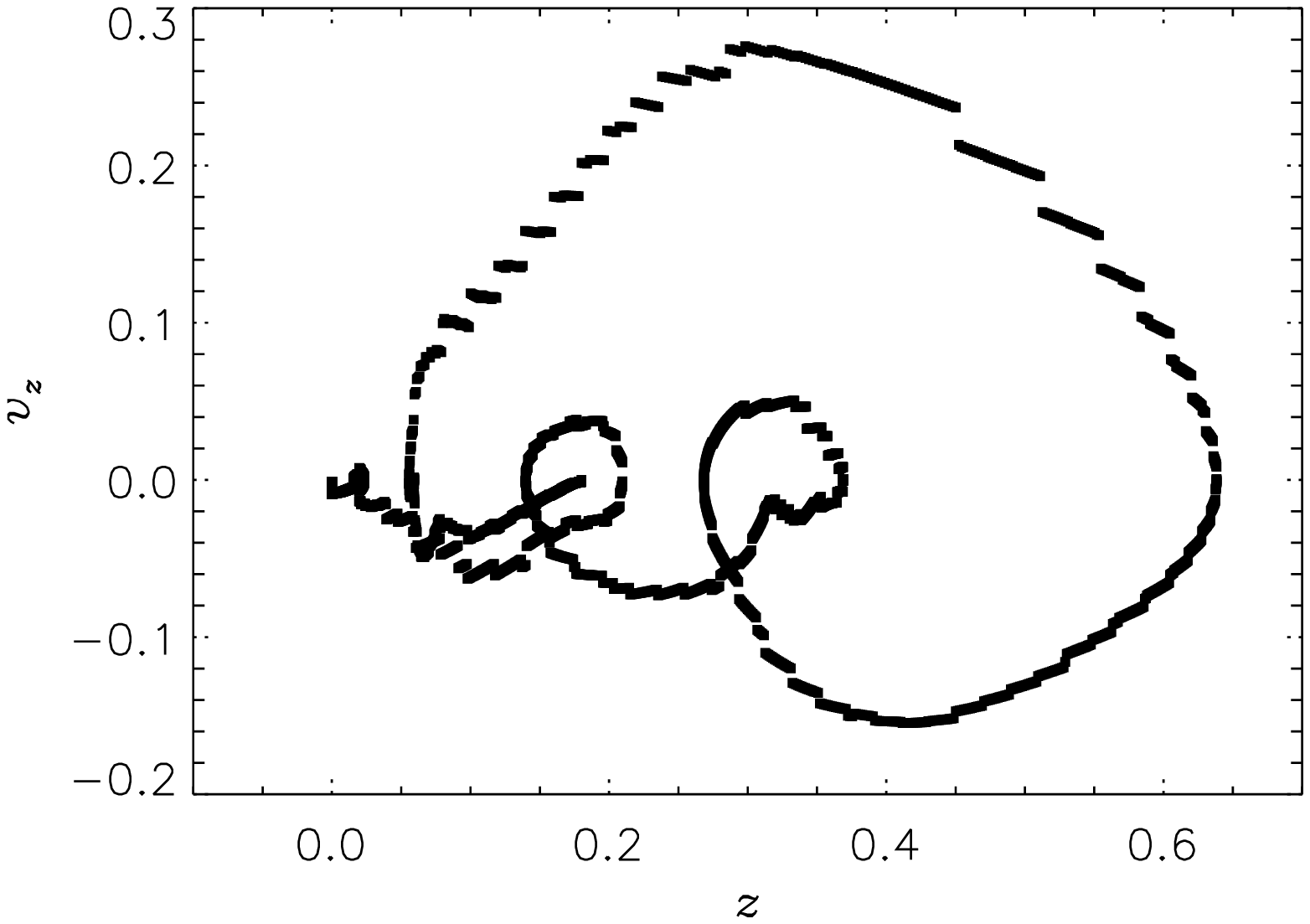}\\
\phantom{.}\hspace{3.45cm}(a)\hspace{6.35cm}(b)\\
 \includegraphics[width=0.5\textwidth,bb=30 15 504 360,clip]{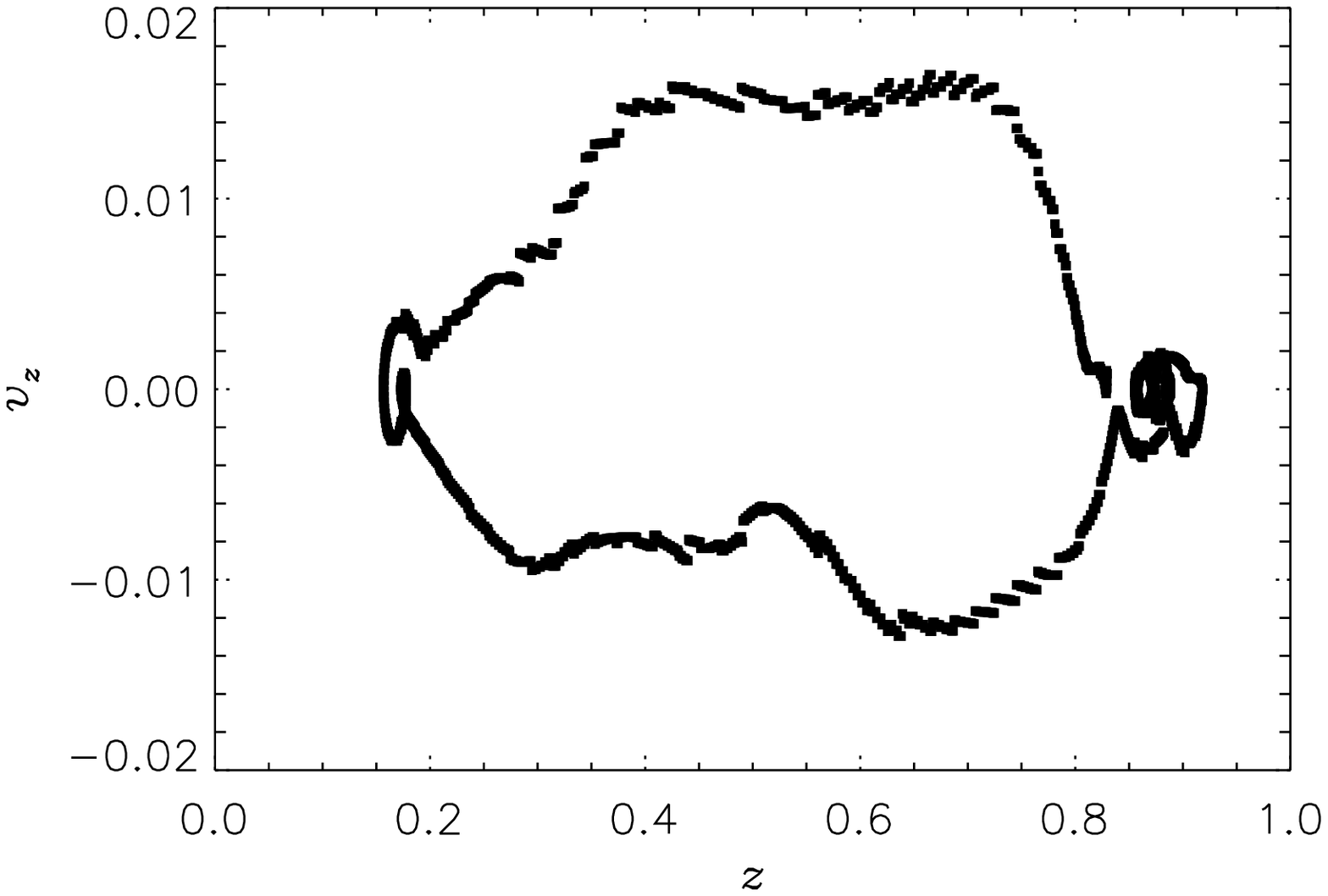}
 \includegraphics[width=0.5\textwidth,bb=30 15 504 360,clip]{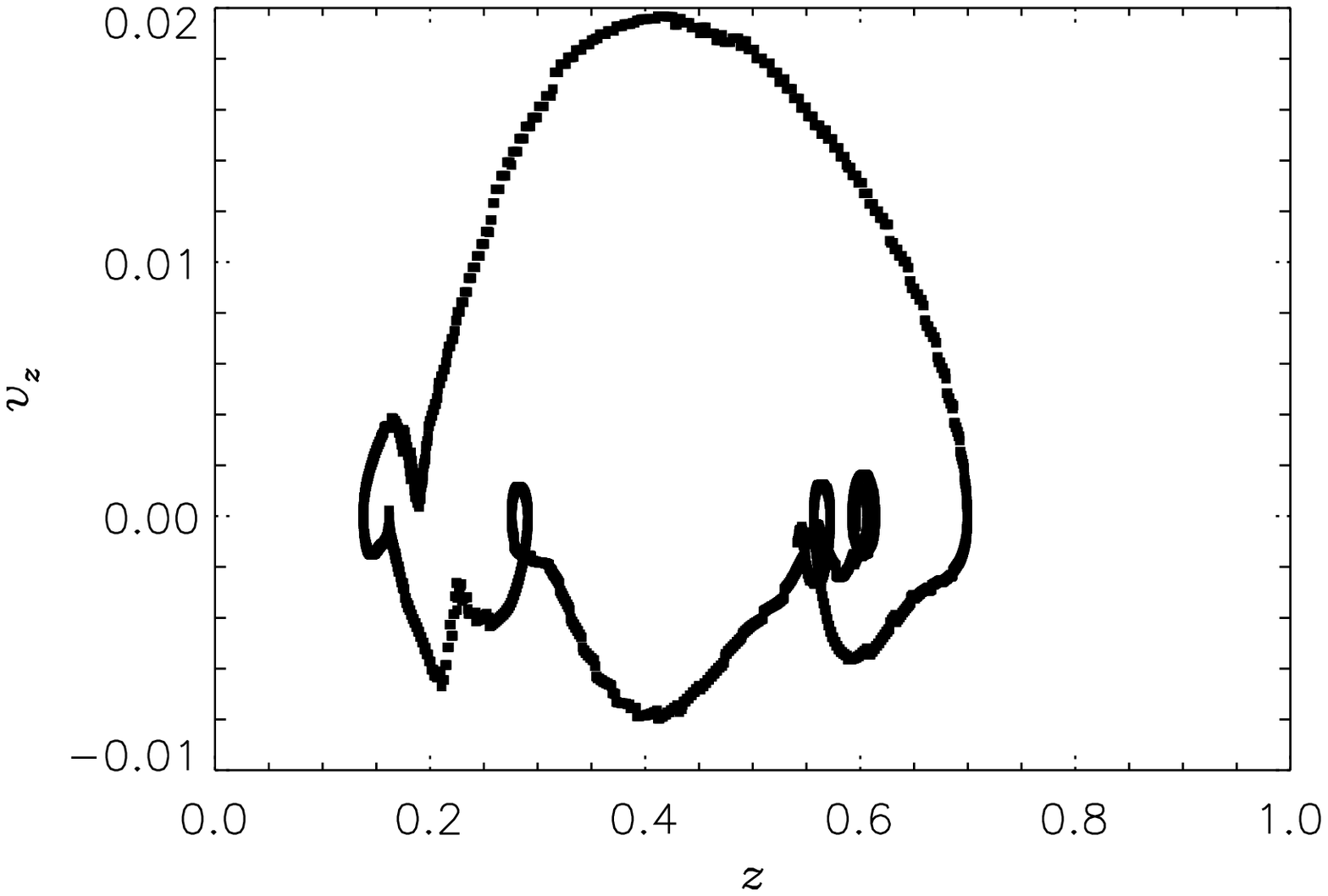}\\
 \phantom{.}\hspace{3.3cm}(c)\hspace{6.35cm}(d)\\
   \caption{Phase trajectories of different fluid particles in the problem
   with a nonmonotonic static temperature profile, $ =0.01$, $b=600$, $n=10$:
   (a, b) $R=20\,000 \approx 1.5R_{\mathrm c}$; {(c, d)
   $R=200\,000 \approx 15R_{\mathrm c}$}.
 }\label{attr_heat}
\end{figure}

Similar analyses were done for the case of the nonmonotonic
static temperature profile. Various features in the obtained
trajectories can be seen in Fig.~\ref{attr_heat}. {The closed
(or nearly closed) loops of the trajectories located near $z=0$
or $z=1$ (Figs~\ref{attr_heat}a, \ref{attr_heat}c,
\ref{attr_heat}d) evidence the presence of small-scale
structures near the lower or the upper layer boundary.} The
trajectories exemplified in Fig.~\ref{attr_heat}b,
\ref{attr_heat}d indicate the presence of small-scale
structures in the bulk of the layer. Thus, the coexistence of
large-scale and small-scale cells can be revealed in the
problem with a nonmonotonic static temperature profile, and the
small-scale structures are distributed over the whole layer
thickness rather than localised in thin sublayers. They move
through the layer, being advected by the large-scale flow. It
should be noted that the spatial trajectories of the particles
have numerous nearly horizontal segments, which is reflected by
the discontinuities of the phase trajectories seen in
Fig.~\ref{attr_heat}.

\section{Conclusion}

We see that the stratification due to the variable thermal
diffusivity, with a sharp kink in the static temperature
profile, can give rise to the development of small-scale flows
superposed onto larger-scale flows, i.e., on the whole, to the
splitting of convection scales. In the case of two rigid
boundaries, small-scale cells are localised in both the upper
and bottom boundary sublayers. However, small-scale cells are
observed only near the upper boundary if it is free. In both
cases, the thickness of the localisation zones of small-scale
cells {does not coincide with the thickness of the sublayer
with the sharp temperature change, $\Delta h$, and declines
with the Rayleigh number increasing. The small-scale structures
become more clear-cut at higher Rayleigh numbers.} Small cells
are advected by large-scale flows; if the upper boundary is
free, this process appears as the sinking of small cells.

{The flows computed for the nonmonotonic profile are much less
ordered than in the case of the rigid or free upper boundary.
They are highly changeable, characterised by permanently
arising descending and ascending plumes, and can on the whole
be assigned to fairly developed turbulence.}

The appearance of phase trajectories in the case of the
nonmonotonic profile also counts in favour of the small-scale
structures being transferred by the large-scale flow. It is
worth noting in this context that, as some observational data
suggest, granules are advected by larger-scale flows in the
solar convection zone. Analyses of correlations between the
brightness variations at two points located not far from each
other suggest that granules may even repeatedly emerge on the
solar surface, playing a relatively passive role in the
convection dynamics \citep{Getling_3}.

Clearly, the simplified models considered here cannot offer a
realistic description of solar convection, the structure of
which depends on a multitude of factors, such as the density
difference across the convection zone, the complex thermal
stratification, the complex equation of state of the matter
(related to the variable ionization degree), radiative heat
exchange, etc. Nevertheless, such models are instructive in
terms of evaluating the possible role of various factors in the
formation of the real pattern.

\section*{Acknowledgement}

We are grateful to the referees for their constructive
criticism and helpful suggestions.


\begin{thebibliography}{00}

\harvarditem{Arakawa}{1966}{Matrix_3}{Arakawa, A.,
Computational design for long-term numerical integration of the
equations of fluid motion:  Two-dimensional incompressible
flow. Part I, J. Comput. Phys., 1, 119-143, 1966; reprinted:
135, 103--114 (CP975697), 1997.}

\harvarditem{Blahut}{1985}{filters}{Blahut, R.E., Fast
Algorithms for Signal Processing, Addison-Wesley, Boston,
1985.}

\harvarditem{Brummell et al.}{1995}{brcattoomre}{Brummell, N.,
Cattaneo, F., \& Toomre, J., Turbulent dynamics in the solar
convection zone, Science, 269, 1370--1379, 1995.}

\harvarditem{Cattaneo et al.}{2001} {catlenzweiss}{Cattaneo,
F., Lenz, D., \& Weiss, N., On the origin of the solar
mesogranulation, Astrophys. J., 563, L91--L94, 2001.}

\harvarditem{DeRosa et al.}{2002}{derosaetal}{DeRosa, M.L.,
Gilman, P.A., \& Toomre, J., Solar multiscale convection and
rotation gradients studied in shallow spherical shells,
Astrophys. J., 581, 1356--1374, 2002.}

\harvarditem{Getling}{1976}{Get76} {Getling, A.V., Convective
motion concentration at the boundaries of a horizontal fluid
layer with inhomogeneous unstable temperature gradient along
the height, Fluid Dyn., 10, 745--750, 1976.}

\harvarditem{Getling}{1980}{Get80}{Getling, A.V., Scales of
convective flows in a horizontal layer with radiative transfer,
Izv., Atmosph. Oceanic Phys., 16, 363--365, 1980.}

\harvarditem{Getling}{1998}{Getling_book}{Getling, A.V.,
Rayleigh--B\'enard Convection: Structures and Dynamics, World
Scientific, Singapore, 1998; URSS, Moscow, 1999.}

\harvarditem{Getling}{2006}{Getling_3} {Getling, A.V., Do
quasi-regular structures really exist in the solar photosphere?
I.~Observational evidence, Solar Phys., 239, 93--111, 2006.}

\harvarditem{Getling et al.}{2013}{gmshch}{Getling, A.V.,
Mazhorova, O.S., \& Shcheritsa, O.V., Concerning the multiscale
structure of solar convection, Geomagn. Aeron., 53, 904--908,
2013.}

\harvarditem{Getling \& Tikhomolov}{2007}{Get2007}{Getling,
A.V., \& Tikhomolov, E.M., Scale splitting in solar convection,
Trudy XI Pulkovskoi mezhdunarodnoi konferentsii po fizike
Solntsa (Proc. 11th Pulkovo Int. Conf. on Solar Physics),
Pulkovo, 2007, pp. 109--112.}

\harvarditem{Kitiashvili et al.}{2012}{kitkosetal}{Kitiashvili,
I.N., Kosovichev, A.G., Mansour, N.N., Lele, S.K., \& Wray,
A.A., Vortex tubes of turbulent solar convection, Phys. Scr.,
86, paper id. 018403, 2012.}

\harvarditem{Kovenya \& Yanenko}{1981}{Yanenko81}{Kovenya,
V.M., \& Yanenko, N.N., The Splitting Method in Problems of Gas
Dynamics, Novosibirsk, Nauka, 1981.}

\harvarditem{Krishan et al.}{2007}{Homology_2}{Krishan, K.,
Kurtuldu, H., Schatz, M.F., Madruga, S., Gameiro, M., \&
Mischaikow, K., Homology and symmetry breaking in
Rayleigh-B\'enard convection: Experiments and simulations,
Phys. Fluids, 19, 17105, 2007.}

\harvarditem{Mazhorova \& Popov}{1980}{Matrix_2} {Mazhorova,
O.S., \& Popov, Yu.P., Methods for the numerical solution of
the Navier--Stokes equations, USSR Comput. Math. Math. Phys.,
20, 202--217, 1980.}

\harvarditem{Mazhorova \& Popov}{1981}{Matrix_REP}{Mazhorova,
O.S., \& Popov, Yu.P., Matrix iteration method of numerical
solution of Navier--Stokes two--dimensional equations, Doklady
Akad. Nauk SSSR, 259, 535--540, 1981.}

\harvarditem{November et al.}{1981}{novetal}{November, L.J.,
Toomre, J., Gebbie, K.B. and Simon, G.W., The detection of
mesogranulation on the Sun, Astrophys. J. Lett., 245,
L123--L126, 1981.}


\harvarditem{Rogers et al.}{2003}{rogersetal}{Rogers, T.M.,
Glatzmaier, G.A., \& Woosley, S.E., Simulations of
two-dimensional turbulent convection in a density-stratified
fluid, Phys. Rev. E, 67, article id. 026315, 2003.}

\harvarditem{Schmalzl et al.}{2004}{schmalzletal}{Schmalzl, J.,
Breuer, M., \& Hansen, U., On the validity of two-dimensional
numerical approaches to time-dependent thermal convection,
Europhys. Lett., 67, 390--396, 2004.}

\harvarditem{Simon \& Leighton}{1964}{simonleighton}{Simon,
G.W., \& Leighton, R.B., Velocity fields in the solar
atmosphere. III. large-scale motions, the chromospheric
network, and magnetic fields, Astrophys. J., 140, 1120--1147,
1964.}

\harvarditem{Stengel et al.}{1982}{stengetal}{Stengel, K.C.,
Oliver, D.S., \& Booker, J.R., Onset of convection in a
variable-viscosity fluid, J. Fluid Mech., 120, 411--431, 1982.}




\end{thebibliography}
\end{document}